\documentstyle[aps,preprint,psfig,tighten]{revtex}

\begin{document}

\title{Optimized Constant Pressure Stochastic Dynamics}
\author{A. Kolb and B. D\"unweg}
\address{Max Planck Institute for Polymer Research,\\
    Ackermannweg 10, D--55128 Mainz, Germany}
\date{\today}
\maketitle

\begin{abstract}
  A recently proposed method for computer simulations in the
  isothermal--isobaric (NPT) ensemble, based on Langevin--type
  equations of motion for the particle coordinates and
  the ``piston'' degree of freedom, is re--derived by
  straightforward application of the standard Kramers--Moyal
  formalism. An integration scheme is developed which
  reduces to a time--reversible symplectic integrator in
  the limit of vanishing friction. This algorithm
  is hence expected to be quite stable for small friction,
  allowing for a large time step. We discuss the optimal choice
  of parameters, and present some numerical test results.
\end{abstract}

\vspace{1cm}

\begin{tabular}[t]{ll}
PACS: & 02.50.-r; 02.70.Ns; 05.20.Gg; 82.20.Fd; 82.20.Wt \\
%
%
Keywords: & Molecular Dynamics; NPT Ensemble; Symplectic Algorithms; \\
          & Stochastic Dynamics; Langevin Equation; 
            Fokker--Planck Equation
\end{tabular}

\section{Introduction}
\label{sec:introduction}

Molecular Dynamics (MD) simulations \cite{Allen:89.1,%
Rapaport:95.1,Frenkel:96.1} are a very efficient tool to study
the statistical properties of thermodynamic systems, especially at high
densities where the acceptance rates of standard Monte Carlo (MC)
simulations \cite{Binder:79.1} are small. They are also very
well suited to study dynamical properties.

MD simulations are most naturally performed within the microcanonical
(NVE) ensemble, while the simple Metropolis MC algorithm leads to the
canonical (NVT) ensemble. However, it is often desirable to study a
system within a different ensemble, and hence methods have been
developed to extend both MC and MD methods to practically every ensemble
of thermodynamics \cite{Allen:89.1,Rapaport:95.1,Frenkel:96.1,%
Binder:79.1,Ferrario:93.1,Sprik:94.1,Sprik:95.1}. For MC methods,
this task is relatively straightforward: One starts from the partition
function or equilibrium probability distribution function for the
pertinent degrees of freedom, and defines a Markov process in the space
of the latter. After verifying the condition of detailed balance, and
making use of ergodicity (which often can be safely assumed, and in some
cases even be shown rigorously), one has established that the process
will ultimately produce fluctuations within the proper equilibrium. For
example, for a simulation in the isothermal--isobaric (NPT) ensemble,
the system is studied in a box with periodic boundary conditions, whose
size is allowed to fluctuate. In order to keep the system homogeneous,
the particle coordinates are instantaneously adjusted to these box
fluctuations, via simple rescaling. We shall here be concerned with the
simplest case only, where the box is just a cube of size $L$ in each
direction. By writing
\begin{equation} \label{boxrescaleqn}
  \vec r_i = L \vec s_i
\end{equation}
one introduces reduced coordinates $\vec s_{i}$ in the unit cube
instead of the original coordinates $\vec r_{i}$ of the
$N$ particles in the box of volume $V = L^{d}$, $d$ denoting
the spatial dimensionality. The abovementioned adjustment of
the particle configuration to the box fluctuations is facilitated
by updating $L$ and the $\vec s_{i}$ independently. The
partition function appropriate to the NPT ensemble is then
\begin{eqnarray} \label{NPTMCpartfunceqn}
  Z & = & \int_{0}^{\infty} d V
        \int d^d \vec r_1 \ldots \int d^d \vec r_N
        \exp \left( - \beta U - \beta P V \right) \nonumber \\
    & = & \int_{0}^{\infty} d V
        \int d^d \vec s_1 \ldots \int d^d \vec s_N V^{N}
        \exp \left( - \beta U - \beta P V \right) ,
\end{eqnarray}
where $U$ denotes the system's potential energy, $P$ is the
external pressure, and $\beta = 1 / (k_{B} T)$. From this
one immediately reads off that one has to run a standard
Metropolis algorithm on the state space $(L, \{\vec s_i\})$,
using an effective Hamiltonian
\begin{equation}
  U_{eff} = U  + P V - N k_B T \ln V .
\end{equation}

The MD approach \cite{Allen:89.1,Rapaport:95.1} to non--microcanonical
ensembles \cite{Allen:89.1,Rapaport:95.1,Frenkel:96.1,%
Ferrario:93.1,Sprik:94.1,Sprik:95.1}, pioneered by
Andersen \cite{Andersen:80.1}, Nos\'e \cite{Nose:84.1}
and Hoover \cite{Hoover:85.1}, is slightly more involved. Like in MC,
one defines an additional dynamical variable whose fluctuations
allow to keep the thermodynamically conjugate variable fixed. 
In our example, this variable is $L$, while $P$ is fixed.
However, the dynamics is not specified via a random
process, but rather by Hamiltonian equations of motion in
the extended space. This requires the definition of
canonically conjugate momenta of the additional variables,
and in turn the introduction of artificial masses, which
have no direct physical meaning but are adjusted in order
to set the time scale of the new fluctuations. The analysis
then proceeds by assuming ergodicity in the extended space,
such that the statistics is described by a microcanonical
ensemble in that space. If the equations of motion have been
chosen properly, then the equilibrium distribution function
of the algorithm, which is obtained by integrating out the
artificial momenta, must coincide with that of the desired
ensemble. For constant--pressure simulations, often the
picture of a ``piston'' is employed. In the last few years many
contributions on extending constant--pressure methods have been
made, treating cases of isotropic boxes as well as non--cubic
cells \cite{Parrinello:80.1,Martyna:92.1,Ciccotti:93.1,Martyna:94.1,%
Procacci:94.1,Feller:95.1}. We will here however be only
concerned with the case of an isotropic box, as in the
original Andersen method \cite{Andersen:80.1} which
produces an isobaric--isenthalpic (NPH) ensemble.

Stochastic dynamics (SD) in its classical form \cite{Schneider:78.1}
can be viewed as a simulation method which is somewhere between
MC and MD, sharing with the former the stochastic element (which also
ensures the ergodicity of the method) and the generation of a
canonical (NVT) ensemble, while being based on continuous
equations of motion, and momenta, like the latter. Instead
of Hamiltonian equations of motion one solves Langevin
equations and adjusts the temperature via the balance
between friction force and stochastic force (fluctuation--dissipation
theorem). It is not surprising that this approach
can be combined with the Andersen method in order to produce
an NPT ensemble. It is the purpose of this paper to show that
the algorithm can be derived in a very simple and straightforward
manner, by exploiting the well--known equivalence of the
Langevin equation with the Fokker--Planck equation \cite{Risken:89.1},
and avoiding the complicated reasoning of a recent publication
\cite{Feller:95.1}. The derivation is outlined in Secs.
\ref{sec:genstochdyn}, \ref{sec:andersen} and \ref{sec:FP-NPT},
where we treat a straightforward generalization of SD to
arbitrary Hamiltonian systems, and apply this to the original
Andersen NPH method, which is thus modified to an NPT method.
We then turn to the question of numerical implementation. Since SD
reduces to standard Hamiltonian dynamics in the limit of zero
friction, and practical applications are often run for rather
small friction, it is useful to use an algorithm which reduces
to a time--reversible symplectic integrator (TRSI) in the
zero--friction limit. It is well--known that TRSIs are
particularly well--suited for Hamiltonian systems
\cite{Sprik:94.1,Sprik:95.1,Procacci:94.1,Tuckerman:92.1,%
Duenweg:97.1}, since, except for roundoff errors, they do
not prefer a particular direction of time and are hence very
stable. Our algorithm is derived and tested in Sec.
\ref{sec:implementation}, where we apply the method to
a simple model system of Lennard--Jones particles, and
pay particular attention to the question how the parameters
should be chosen for optimum performance. Finally, we conclude
in Sec. \ref{sec:conclusion}.

\section{Generalized Stochastic Dynamics}
\label{sec:genstochdyn}

Our starting point is a set of generalized coordinates $q_{i}$ and
canonically conjugate momenta $p_{i}$, such that the Hamiltonian
equations of motion read
\begin{equation} \label{generalhamiltoneq}
\dot{q}_{i} =   \frac{\partial {\cal H}}{\partial p_{i}} \hspace{2cm}
\dot{p}_{i} = - \frac{\partial {\cal H}}{\partial q_{i}},
\end{equation}
where ${\cal H}$ is the Hamilton function. These equations of motion are
generalized to their stochastic versions by adding friction forces
with damping parameters $\gamma_{i} \left( \left\{ q_{i} \right\} \right)$
and stochastic forces with noise strengths
$\sigma_{i} \left( \left\{ q_{i} \right\} \right)$ (note that a dependence
on the generalized coordinates is permitted):
\begin{equation} \label{generallangevin}
\dot{q}_{i} = \frac{\partial {\cal H}}{\partial p_{i}} \hspace{2cm}
\dot{p}_{i} = - \frac{\partial {\cal H}}{\partial q_{i}}
              - \gamma_{i} \frac{\partial {\cal H}}{\partial p_{i}}
              + \sigma_{i} \eta_{i} (t),
\end{equation}
where the Gaussian white noise $\eta_{i}$ satisfies the usual relations
\begin{equation}
\left< \eta_{i} (t) \right> = 0 \hspace{2cm}
\left< \eta_{i} (t) \eta_{j} (t^{\prime}) \right> =
2 \delta_{ij} \delta \left( t - t^{\prime} \right).
\end{equation}

Equivalently, the stochastic process is described by the Fokker--Planck
equation \cite{Risken:89.1} which describes the time evolution of the
probability distribution function $\Phi$ in the full space of stochastic
variables. For an arbitrary set of variables $x_{\nu}$ (in our case both
$q_{i}$ and $p_{i}$) it reads
\begin{equation} \label{eq:FP_formal}
  {\partial \Phi \over \partial t} = i \hat{L}_{FP} \Phi =
  - \sum_{\nu} {\partial \over \partial x_{\nu}} D_{\nu}^{(1)} \Phi
  + \sum_{\mu \nu} {\partial^{2} \over \partial x_{\mu} \partial x_{\nu} }
    D_{\mu \nu}^{(2)} \Phi,
\end{equation} 
where the right hand side of the equation defines the Fokker--Planck
operator $\hat{L}_{FP}$. Drift and diffusion coefficients $D_{\nu}^{(1)}$
and $D_{\mu \nu}^{(2)}$ are related to the short--time behavior of the
process via the Kramers--Moyal expansion \cite{Risken:89.1}:
\begin{eqnarray} \label{eq:KramersMoyal}
  D_{\nu}^{(1)} & = & \lim_{\Delta t\to 0} {1\over \Delta t}
                      \langle\Delta x_{\nu} (\Delta t) \rangle 
                    = \lim_{\Delta t \to 0}{1\over \Delta t}
      \langle x_{\nu} (t + \Delta t) - x_{\nu} (t) \rangle \nonumber \\
  D_{\mu \nu}^{(2)} & = & \lim_{\Delta t\to 0}{1 \over 2 \Delta t}
                          \langle \Delta x_{\mu} (\Delta t)
                                  \Delta x_{\nu} (\Delta t) \rangle .
\end{eqnarray}
Straightforward evaluation of these moments for the present case yields
directly the Fokker--Planck operator:
\begin{eqnarray} \label{eq:FP-operator} \nonumber
i \hat{L}_{FP} & = & - \sum_{i} \frac{\partial}{\partial q_{i}}
                              \frac{\partial {\cal H}}{\partial p_{i}}
                   - \sum_{i} \frac{\partial}{\partial p_{i}}
                     \left( - \frac{\partial {\cal H}}{\partial q_{i}}
                            - \gamma_{i} 
                              \frac{\partial {\cal H}}{\partial p_{i}}
                          \right)
                   + \sum_{i} \frac{\partial^{2}}{\partial p_{i}^{2}}
                              \sigma_{i}^{2}  \nonumber \\
             & = & - \sum_{i} \frac{\partial {\cal H}}{\partial p_{i}}
                              \frac{\partial}{\partial q_{i}}
                   + \sum_{i} \frac{\partial {\cal H}}{\partial q_{i}}
                              \frac{\partial}{\partial p_{i}}
                   + \sum_{i} \frac{\partial}{\partial p_{i}}
                     \left( \gamma_{i} 
                            \frac{\partial {\cal H}}{\partial p_{i}}
                            + \sigma_{i}^{2}
                            \frac{\partial}{\partial p_{i}}
                            \right) .
\end{eqnarray}

A canonical ensemble is generated if the Boltzmann distribution
is the stationary distribution of the process (due to the stochastic
element, the process is usually ergodic, such that only one
stationary distribution exists). For the present case one finds
\begin{equation} \label{eq:stationarydistrib}
i \hat{L}_{FP} \exp \left( - \beta {\cal H} \right) =
\sum_{i} \frac{\partial}{\partial p_{i}} \left( \gamma_{i}
- \beta \sigma_{i}^{2} \right) \frac{\partial {\cal H}}{\partial p_{i}}
\exp \left( - \beta {\cal H} \right),
\end{equation}
which vanishes if
\begin{equation} \label{eq:fluctdiss}
\sigma_{i}^{2} = k_{B} T \gamma_{i} .
\end{equation}
This relation is the generalized fluctuation--dissipation theorem which
friction and noise have to satisfy in order to generate a canonical ensemble.

\section{Andersen Extended System}
\label{sec:andersen}

The Andersen method \cite{Andersen:80.1} uses the box volume $V = L^{d}$ and
the scaled coordinates $\vec s_{i}$, see Eqn. \ref{boxrescaleqn}, as degrees
of freedom. For the particle velocities one obtains
\begin{equation} \label{coordtimederiveqn}
  \dot{\vec r_{i}} = L \dot{\vec s_{i}} + \dot{L} \vec s_{i},
\end{equation}
however, the second term is deliberately omitted in order to achieve
independent fluctuations of $L$ and $\vec s_{i}$. Hence the method
amounts to {\em postulating} the Lagrangian
\begin{eqnarray} \label{eq:lagrangian}
  {\cal L} & = & \sum_{i} {L^{2} \over 2} m_{i} \dot{ \vec s_i }^2
               - \sum_{i < j} v_{ij} \left( L, \{ \vec s_{i} \} \right)
               + {Q \over 2} \dot{V}^{2} - P V,
\end{eqnarray}
where $m_{i}$ denotes the mass of the $i$th particle, $Q$ is the artificial
piston mass or box mass, and $v_{ij}$ is the interaction potential between
particles $i$ and $j$ (the generalization to three-- and many--body forces
is straightforward). The Hamiltonian is then obtained via Legendre
transformation
\begin{equation} \label{eq:hamiltonian}
  {\cal H}  =  \sum_{i} \vec \pi_{i} \dot{\vec s_{i}}
               + \Pi_{V} \dot{V} - {\cal L}
            =  \sum_{i} {1 \over 2 L^{2} m_{i}} \vec \pi_{i}^{2}
               + \sum_{i < j} v_{ij} 
               + {1 \over 2 Q} \Pi_{V}^{2} + P V,
\end{equation}
where we have used the canonically conjugate momenta
\begin{equation} \label{eq:canonmomenta}
  \Pi_{V}      = \frac{\partial {\cal L}}{\partial \dot{V}}
               = Q \dot{V}, \hspace{2cm}
  \vec \pi_{i} = \frac{\partial {\cal L}}{\partial \dot{\vec s_i}}
               = m_{i} L^{2} \dot{\vec s_{i}},
\end{equation}
such that the Hamiltonian equations of motion read
\begin{eqnarray} \label{eq:hamiltonian_motion}
  \dot{\vec s_{i}} = {1 \over L^{2} m_{i}} \vec \pi_{i} & \hspace{1cm} &
  \dot{\vec \pi_{i}} = L \vec f_{i} \\ \nonumber
  \dot{V} = {1 \over Q} \Pi_{V}  & \hspace{1cm} & 
  \dot{\Pi}_{V} = {\cal P} - P,
\end{eqnarray}
where $\vec f_{i}$ is the force acting on the $i$th particle, and
${\cal P}$ abbreviates the ``instantaneous'' pressure
\begin{equation} \label{eq:instant_pressure}
  {\cal P} = {L \over d V} \sum_{i < j} {\vec f}_{ij} \vec s_{ij}
  + {1 \over d L^{2} V} \sum_{i}{1 \over m_i} \vec \pi_{i}^{2} ,
\end{equation}
$\vec f_{ij}$ being the force acting between particle $i$ and $j$,
while $\vec s_{ij} = \vec s_{i} - \vec s_{j}$.

Obviously, $\cal H$ is a constant of motion. Apart from the kinetic
energy of the piston, and the deviation of the simulated
molecular kinetic energy from the true kinetic energy, which
are both small corrections \cite{Andersen:80.1},
$\cal H$ is just the enthalpy. For this reason, the
method produces the NPH ensemble.

\section{Langevin Equation for the NPT Ensemble}
\label{sec:FP-NPT}
 
The idea of Feller {\em et al.} \cite{Feller:95.1} was to replace the
canonical equations of motion (Eqn. \ref{eq:hamiltonian_motion})
by a Langevin stochastic process. It was designed to avoid
oscillations of the box volume (``ringing'' of the box). In
Ref.~\cite{Feller:95.1} an infinite set of harmonic oscillators
coupled to the box piston was used to prove the correctness
of the approach. However, since the original Andersen method is
based on a Hamiltonian system, the results which have been derived in Sec.
\ref{sec:genstochdyn} can be directly applied. Therefore the stochastic
equations of motion read
\begin{eqnarray} \label{eq:langevin_motion}
  \dot{\vec \pi_{i}} & = & L \vec f_{i}
                           - \frac{\gamma_{i}}{L^{2} m_{i}} \vec \pi_{i}
                         + \sqrt{k_{B} T \gamma_{i}}
                           \vec \eta_{i} (t) \\ \nonumber
  \dot{\Pi}_{V} & = & {\cal P} - P
                      - \frac{\gamma_{V}}{Q} \Pi_{V}
                      + \sqrt{k_{B} T \gamma_{V}}
                        \vec \eta_{V} (t) \\ \nonumber
  \langle \eta_{i}^{\alpha} \rangle = 
  \langle \eta_{V} \rangle & = & 0 \\ \nonumber
  \langle \eta_{i}^{\alpha} (t) \eta_{j}^{\beta} (t^{\prime}) \rangle & = &
  2 \delta_{ij} \delta_{\alpha \beta} \delta(t - t^{\prime})
  \\ \nonumber
  \langle \eta_{V} (t) \eta_{V} (t^{\prime}) \rangle & = &
  2 \delta(t - t^{\prime}) \\ \nonumber
  \langle \eta_{i}^{\alpha} (t) \eta_{V} (t^{\prime}) \rangle & = & 0,
\end{eqnarray}
while the equations of motion for $\vec s_{i}$ and $V$ remain unchanged.
Here $\alpha$ and $\beta$ denote Cartesian directions, and the
fluctuation--dissipation relation has already been taken into account.

This method generates the correct NPT ensemble, as is seem from
writing down the partition function which naturally arises from
the algorithm (cf. Sec. \ref{sec:genstochdyn}):
\begin{equation} \label{eq:nptpartfunc}
Z = \int d \Pi_V \int d^d \vec \pi_1 \ldots \int d^d \vec \pi_N
    \int d V \int d^d \vec s_1 \ldots \int d^d \vec s_N \quad
    \exp \left( - \beta {\cal H} \right) ,
\end{equation}
where one obviously has to use the Hamiltonian of the Andersen
method, see Eqn. \ref{eq:hamiltonian}. Integrating out the
momenta, one sees directly that this is, apart from unimportant
prefactors, identical to Eqn. \ref{NPTMCpartfunceqn}, i.~e.
the correct partition function of the NPT ensemble.

Note that there is still considerable freedom in the choice of the
friction parameters. Feller {\em et al.} chose $\gamma_i = 0$
and $\gamma_V = \mbox{\rm const.}$. This is tantamount to coupling
{\em only} the piston degree of freedom to the heat bath.
Since this degree of freedom is tightly coupled to all the others,
the method produces the same NPT ensemble as the more general case
$\gamma_i \neq 0$. However, we view this latter case, which 
also includes a direct coupling of the particles to the heat bath,
as more advantageous, not for fundamental reasons, but rather
for practical ones: As in standard SD, every degree of freedom
is thermostatted {\em individually}. For this reason, local
instabilities, arising from discretization errors, are efficiently
corrected for, without spreading throughout the system. Loosely
spoken, a particle which happens to be too ``hot'' will be ``cooled
down'' by its local friction, while in the opposite case it will
be ``heated up'' by the noise. For this reason, the SD version
with $\gamma_i \neq 0$ allows for a slightly larger time step than
the pure Andersen method, while the Feller {\em et al.} version
($\gamma_i = 0$, $\gamma_V \neq 0$) is not more stable, involving
only {\em global} thermostatting. We have chosen
\begin{equation} \label{ourfriction}
\gamma_i = \gamma_0 L^2,
\end{equation}
while choosing a constant value for $\gamma_V$. Then the
Langevin equation for $\vec \pi_i$ is rewritten as
\begin{equation} \label{ourlangevin}
\dot{\vec \pi_i} = L \left( \vec f_i - \frac{\gamma_0}{m_i}
                   \frac{\vec \pi_i}{L} + \sqrt{k_B T \gamma_0}
                   \vec \eta_i(t) \right) ;
\end{equation}
this fits naturally to standard SD, where the friction
force is $- \gamma_0 \vec p_i / m_i = - \gamma_0 \vec \pi_i / (L m_i)$.
Further details on the implementation
will be given in Sec. \ref{sec:implementation}; there we will
also discuss the choice of the parameters $\gamma_0$
and $\gamma_V$, as well as of the piston mass $Q$,
which are all irrelevant for the statistical properties
of the system, but of great importance for the equilibration
properties of the algorithm.

\section{Implementation}
\label{sec:implementation}

\subsection{Symplectic Integrator}
\label{sec:algorithm}

Symplectic time--reversible integrators are known to be extremely useful
for molecular dynamics simulations of Hamiltonian systems
\cite{Sprik:94.1,Sprik:95.1,Procacci:94.1,Tuckerman:92.1,Duenweg:97.1,%
Tuckerman:94.1,Tuckerman:94.2}.
This is so because, except for roundoff errors, they conserve
the phase--space volume exactly (note the intimate relation to entropy
production), and do not mark any particular direction of time
(note that a global drift in the algorithm breaks this time--reversal
symmetry). Hence they are very stable and allow for a large time step.
The most common example (actually, the lowest--order scheme)
is the well--known Verlet algorithm in its various formulations
\cite{Allen:89.1}.

A particularly transparent way to construct these algorithms
\cite{Tuckerman:92.1} is based on the Liouville operator
$i \hat L$, which is just the Fokker--Planck operator
in the special case of vanishing friction (see Eqn. \ref{eq:FP-operator}).
Noticing that the formal solution of Eqn. \ref{eq:FP_formal} is just
\begin{equation} \label{formalsolution}
\Phi(t) = \exp \left( i \hat L t \right) \Phi(0),
\end{equation}
one focuses directly on the operator $\exp \left( i \hat L t \right)$,
and decomposes $\hat L$ into a sum of simpler operators,
\begin{equation} \label{sumofsimpleroperators}
i \hat{L} = \sum_{k=1}^{n} i \hat{L}_k,
\end{equation}
where the imaginary unit is extracted for convenience;
$\hat{L}$ as well as each of the $\hat L_k$ is self--adjoint.
Now, the exact time development within a time step
$\Delta t$ is replaced by an approximate one by
using the factorization
\begin{equation} \label{eq:factorapprox}
e^{i \hat L \Delta t} \to
e^{i \hat L_1 \Delta t / 2}
e^{i \hat L_2 \Delta t / 2} \ldots
e^{i \hat L_{n-1} \Delta t / 2}
e^{i \hat L_{n} \Delta t}
e^{i \hat L_{n-1} \Delta t / 2} \ldots
e^{i \hat L_2 \Delta t / 2}
e^{i \hat L_1 \Delta t / 2}.
\end{equation}
This scheme is automatically phase--space conserving, since
each of the operators is unitary, and time--reversible
symmetric, since the inverse operator is just the original
operator, evaluated for $- \Delta t$. Now, if each of the
operators $\hat L_k$ is simple enough, the action of $e^{i \hat L_k t}$
on a phase space point can be calculated trivially, such that the
algorithm is a succession of simple updating steps.

Specifically, for the Andersen equations of motion (see Sec.
\ref{sec:andersen}) we choose the operators
\begin{eqnarray} \label{ands_liouv}
  i \hat{L}_{1} & = & - \sum_i L \vec f_i
  \frac{\partial}{\partial \vec \pi_i} \\
  i \hat{L}_{2} & = & - \left( {\cal P} - P \right)
  \frac{\partial}{\partial \Pi_V} \nonumber \\
  i \hat{L}_{3} & = & - \frac{\Pi_V}{Q}
  \frac{\partial}{\partial V} \nonumber \\
  i \hat{L}_{4} & = & - \sum_i \frac{\vec \pi_i}{L^2 m_i}
  \frac{\partial}{\partial \vec s_i} , \nonumber
\end{eqnarray}
resulting in the following updating scheme:
\begin{enumerate}
\item  $ \vec \pi_i(t) \rightarrow \vec \pi_i(t + \Delta t / 2)
       = \vec \pi_i(t) + L(t) \vec f_i (t) \Delta t / 2 $

\item  $ \Pi_{V}(t) \rightarrow \Pi_{V} (t + \Delta t / 2)
       = \Pi_{V}(t) + \left( {\cal P} - P \right) \Delta t / 2 $

       (note that for the evaluation of ${\cal P}$, one has
       to take the old positions $\vec s_i(t)$ and the old
       box size $L(t)$, but already the updated
       momenta $\vec \pi_i(t + \Delta t / 2)$)

\item  $ V(t) \rightarrow V(t + \Delta t / 2)
       = V(t) + Q^{-1} \Pi_{V} (t + \Delta t / 2) \Delta t / 2 $

\item  $ \vec s_{i}(t) \rightarrow \vec s_{i} (t + \Delta t)
       = \vec s_{i}(t) +
     \frac{\vec \pi_i(t + \Delta t / 2)}{L^2(t + \Delta t / 2) m_{i}}
     \Delta t $

\item  $ V(t + \Delta t / 2) \rightarrow V(t + \Delta t)
       = V(t + \Delta t / 2) +
         Q^{-1} \Pi_{V}(t + \Delta t / 2) \Delta t / 2$

\item  $ \Pi_{V} (t + \Delta t / 2) \rightarrow \Pi_{V} (t + \Delta t)
       = \Pi_{V} (t + \Delta t / 2) 
       + \left({\cal P} - P \right) \Delta t / 2$

       (for evaluation of ${\cal P}$, one uses $\vec s_i(t + \Delta t)$,
       $L(t + \Delta t)$, and $\vec \pi_i(t + \Delta t / 2)$)

\item  $ \vec \pi_i(t + \Delta t / 2) \rightarrow 
         \vec \pi_i(t + \Delta t) = 
         \vec \pi_i(t + \Delta t / 2) 
         + L(t + \Delta t) \vec f_{i} (t + \Delta t) \Delta t / 2$ .
\end{enumerate}

It is often convenient to formulate the algorithm in terms
of the conventional variables
\begin{equation} \label{unscaledvariables}
\vec r_i (t) = L(t) \vec s_i (t)  \hspace{3cm}
\vec p_i (t) = L(t)^{-1} \vec \pi_i (t),
\end{equation}
which are however not canonically conjugate with respect
to each other. The pressure, in terms of these variables,
is written as
\begin{equation} \label{pressureunscaled}
  {\cal P} = {1 \over d V} \sum_{i < j} {\vec f}_{ij} \vec r_{ij}
  + {1 \over d V} \sum_{i}{1 \over m_i} \vec p_{i}^{\,2} ,
\end{equation}
and the updating scheme, which now involves various
rescaling steps, proceeds as follows:
\begin{enumerate}
\item $ \vec p_i^{\,\prime} = \vec p_i (t) + \vec f_i (t)
        \Delta t / 2$

\item ${\cal P}$ is evaluated using Eqn. \ref{pressureunscaled}
      with $\vec r_i (t)$, $L (t)$, and $\vec p_i^{\,\prime}$;
      then $\Pi_V$ is updated as before:

      $ \Pi_{V} (t + \Delta t / 2)
      = \Pi_{V} (t) + \left( {\cal P} - P \right) \Delta t / 2 $      

\item $ V(t + \Delta t / 2) = V(t) +
        Q^{-1} \Pi_{V} (t + \Delta t / 2) \Delta t / 2 $

\item $ \vec r_i^{\,\prime} = \vec r_i (t) +
      \frac{L^2 (t)}{L^2 (t + \Delta t / 2)}
      \frac{\vec p_i^{\,\prime}}{m_i} \Delta t $

\item $ V(t + \Delta t) = V(t + \Delta t / 2) +
        Q^{-1} \Pi_{V}(t + \Delta t / 2) \Delta t / 2$

      followed by two rescaling steps:
      \begin{itemize}
      \item $ \vec r_i (t + \Delta t) = 
            \frac{L(t + \Delta t)}{L(t)} \vec r_i^{\,\prime} $
      \item $ \vec p_i^{\,\prime \prime} =
            \frac{L(t)}{L(t + \Delta t)} \vec p_i^{\,\prime} $
      \end{itemize}

\item ${\cal P}$ is evaluated using Eqn. \ref{pressureunscaled}
      with $\vec r_i(t + \Delta t)$, $L(t + \Delta t)$, and
      $\vec p_i^{\,\prime \prime}$; then

      $ \Pi_{V} (t + \Delta t) = \Pi_{V} (t + \Delta t / 2) 
        + \left({\cal P} - P \right) \Delta t / 2$
      
\item $ \vec p_i (t + \Delta t) = \vec p_i^{\,\prime \prime} +
        \vec f_i (t + \Delta t) \Delta t / 2 $ .
\end{enumerate}

So far the algorithm has been developed for the case without
friction and noise. For the case with friction and noise, we
simply use the scheme given above, and introduce the following
replacements:
\begin{eqnarray} \label{langevinreplacements}
\vec f_i \Delta t / 2 & \rightarrow &
\vec f_i \Delta t / 2
- \gamma_0 \frac{\vec p_i}{m_i} \Delta t / 2
+ \sqrt{k_B T \gamma_0 \Delta t} \vec z_i \\ \nonumber
\left( {\cal P} - P \right) \Delta t / 2 & \rightarrow &
\left( {\cal P} - P \right) \Delta t / 2
- \gamma_V \frac{\Pi_V}{Q} \Delta t / 2
+ \sqrt{k_B T \gamma_V \Delta t} z_V .
\end{eqnarray}
Here $z_i^{\alpha}$ and $z_V$ denote uncorrelated random numbers with
zero mean and unit variance; for simplicity we sample them from
a uniform distribution via
\begin{equation} \label{fromuniform}
z = \sqrt{12} \left(u - \frac{1}{2} \right)
\end{equation}
where $u$ is uniformly distributed on the unit interval.
The momenta which occur in Eqn. \ref{langevinreplacements}
are, for simplicity, taken as $\vec p_i (t)$ in step (1),
$\Pi_V (t)$ in step (2), $\Pi_V (t + \Delta t / 2)$ in
step (6), and $\vec p_i^{\,\prime \prime}$ in step (7).

\subsection{Choice of Parameters}
\label{sec:choice_parameters}

We start from the observation that a molecular system is characterized by a
typical molecular frequency $\omega_0$, which can be viewed as the
``Einstein'' frequency of oscillations of an atom in its ``cage''
\cite{Hansen:86.1}. With use of the intermolecular potential $v(\vec{r})$ one
gets
\begin{eqnarray} \label{eq:einsteinfrequency}
  \omega_0^2 = {\rho \over d m} 
  \int d^{d} \vec r g(\vec{r}) \nabla^2 v(\vec{r}),
\end{eqnarray}
with $\rho$ being the particle number density, $m$ the mass of the molecules,
and $g(\vec{r})$ the pair distribution function. Alternatively, one can define
a molecular time scale by the time which a sound wave needs for traveling the
nearest neighbor distance \cite{Nose:84.1}. However, both frequencies coincide
by order of magnitude. This frequency governs the time step $\Delta t$ which
one has to choose in order to keep the MD algorithm stable; a typical rule of
thumb says $\Delta t = (1/50) (2 \pi / \omega_0)$.

Similarly, the piston degree of freedom performs oscillations, if it is
simulated with very weak friction in the NVT ensemble. Following Nos\'e
\cite{Nose:84.1}, we can estimate their frequency $\Omega_0$ quite easily.
Within a linearized approximation, the isothermal compressibility
\begin{equation} \label{eq:compressibility}
  \kappa_{T} = - \frac{1}{V} \frac{\partial V}{\partial P} =
  \frac{1}{V k_B T} 
  \left( \left< V^2 \right> - \left< V \right>^2 \right)  
\end{equation}
controls the relation between pressure fluctuations $\delta {\cal P} = {\cal
  P} - P$ and volume fluctuations $\delta V = V - \left< V \right>$ via
\begin{equation} \label{eq:piston_fluctuations}
  \delta {\cal P} = \frac{\partial P}{\partial V} \delta V
  = - \frac{1}{V \kappa_T} \delta V .
\end{equation}
Therefore, one concludes from Eqn. \ref{eq:hamiltonian_motion} by trivial
insertion
\begin{equation} \label{eq:box_oscillations}
  \frac{d^2}{dt^2} \delta V = 
  - \frac{1}{Q V \kappa_T} \delta V ,
\end{equation}
which is the equation of motion of a harmonic oscillator with frequency
\begin{equation} \label{eq:box_frequency}
  \Omega_0^2 = \frac{1}{Q V \kappa_T} .
\end{equation}
Obviously, the piston mass has to be chosen small enough such that the system
can adjust its volume sufficiently fast. On the other hand, it cannot be
chosen too small, since then $\Omega_0$ becomes too large, see Eqn.
\ref{eq:box_frequency}. Clearly, one does not want $\Omega_0$ to exceed
$\omega_0$, since otherwise the simulation would need an unnecessarily small
time step. The optimum piston mass is thus found from the resonance condition
$\Omega_0 = \omega_0$ \cite{Nose:84.1}, which yields a quite different value
for $Q$ (by a factor of $L^{-2}$) than Andersen's original suggestion
\cite{Andersen:80.1} --- this original criterion has turned out to be not
correct. The similar frequencies of the molecular oscillator and the box
volume lead to a very quick energy transfer between them, resulting in a very
efficient equilibration. However, one will often choose a substantially larger
value for $Q$ in order to separate the time scales, such that the molecular
motion on short length and time scales is largely unaffected by the piston
motion. Regardless of the precise choice for $Q$, one should note that keeping
$\Omega_0$ constant implies a scaling of $Q$ with the inverse system size.

When the coupling to the heat bath with friction and noise is added, the
question arises how to choose the damping parameters $\gamma_0$ and
$\gamma_V$. Let us hence study a (deterministic) damped harmonic oscillator,
\begin{equation} \label{eq:damped_oscillator}
  m \ddot{x} + \gamma \dot{x} + m \omega_0^2 x = 0 .
\end{equation}
Obviously, for small $\gamma$ the damping can practically be neglected. On the
other hand, for $\gamma \approx m \omega_0$, damping force and harmonic force
are of the same order of magnitude (for a harmonic oscillator, we can estimate
the velocity via $\dot{x} \approx \omega_0 x$). The exact calculation yields
\begin{equation} \label{eq:aperiodic}
  \gamma_c = 2 m \omega_0
\end{equation}
for the ``aperiodic limit''. This value quantifies the qualitative distinction
between ``weak'' and ``strong'' damping, denoting the boundary between
oscillatory behavior ($\gamma < \gamma_c$) and pure relaxational dynamics
($\gamma > \gamma_c$). Only in the weak damping case $\gamma \ll \gamma_c$,
the fastest time scale (which governs the time step) is given by $\omega_0$,
while for $\gamma \gg \gamma_c$ the damping term dominates, requiring a
smaller time step. For this reason, $\gamma$ values beyond $\gamma_c$ are
clearly undesirable. Even worse, for $\gamma \gg \gamma_{c}$ the relaxation
contains also a very slow component, whose characteristic time is, for large
$\gamma$, given by $\gamma / (m \omega_{0}^{2})$. Therefore, $\gamma =
\gamma_{c}$ is clearly the optimum value for fast equilibration.

However, for the single--particle damping $\gamma_0$ we typically choose a
value which is between one and two orders of magnitude smaller than
$\gamma_c$. This is in accord with the philosophy of simulating the system
rather close to its true Hamiltonian dynamics, such that at least on the local
scales, both spatially and temporally, the dynamics can be considered as
realistic. We hence use the coupling to the heat bath mainly for additional
stabilization, deliberately keeping the molecular oscillations in the
simulation.

On the other hand, there is no analogous argument of ``physical realism''
for the box motion, which is intrinsically unphysical. One is therefore
clearly led to the choice $\gamma_{V} = \gamma_{Vc}$, such that the ringing
is just avoided, while the box motions are still sufficiently fast. The same
conclusion was obtained by Feller {\em et al.} \cite{Feller:95.1}, where
however no theoretical background was provided.

Using this choice, one is left with only {\em one} time scale for the
piston motion, given by $\Omega_{0}$, and this is in turn adjusted according
to the needs of the simulation: If one is only interested in statics,
then ``resonant'' coupling is desirable (i.~e. $\Omega_{0} \approx
\omega_{0}$, and also $\gamma_{0} \approx \gamma_{c}$), at the expense
of distorting the motions even on the molecular scale. If, on the other
hand, it is desired to realistically simulate the molecular oscillations,
one should enforce a separation of time scales by choosing both
$\Omega_{0} \ll \omega_{0}$ and $\gamma_{0} \ll \gamma_{c}$.

\subsection{Numerical Test}
\label{sec:test}

We study a system containing 100 particles interacting via a
truncated Lennard--Jones potential whose attractive
part is cut off:
\begin{equation} \label{eq:LJ-potential}
  U_{LJ} (r) = 4 \epsilon \left[
  \left( {\sigma \over r} \right)^{12} -
  \left( {\sigma \over r} \right)^{6} +
  {1 \over 4} \right];  \hspace{0.5cm} r < 2^{1/6} \sigma. 
\end{equation}
We choose Lennard--Jones units where the parameters $\sigma$ and
$\epsilon$ as well as the particle mass $m$ are set to unity.

\begin{figure}
  \centerline{\psfig{figure=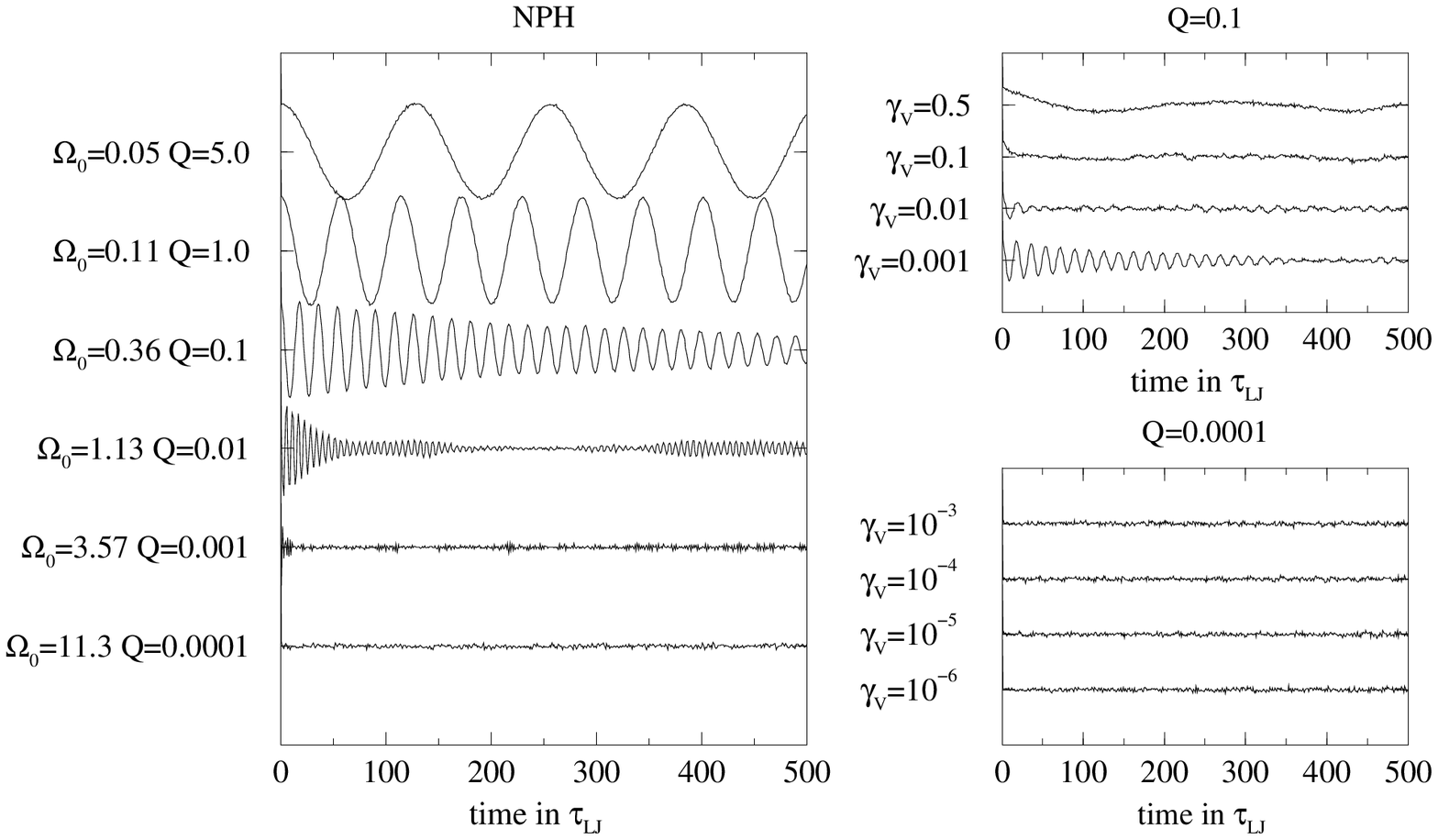,width=14cm}}
  \caption{
    The left hand side shows the autocorrelation function of the instantaneous
    pressure fluctuations using the Andersen algorithm without stochastic
    forces. Various values of $Q$ are used as indicated in the figure. On the
    right hand side the same function is displayed for stochastic dynamics,
    using the box masses $Q = 0.1$ and $Q = 0.0001$ and various damping
    constants $\gamma_V$ as indicated in the figure (molecular damping
    $\gamma_{0} = 0$ in all cases).  }
  \label{fig:pr-autocorrelation}
\end{figure}

\begin{figure}
  \centerline{\psfig{figure=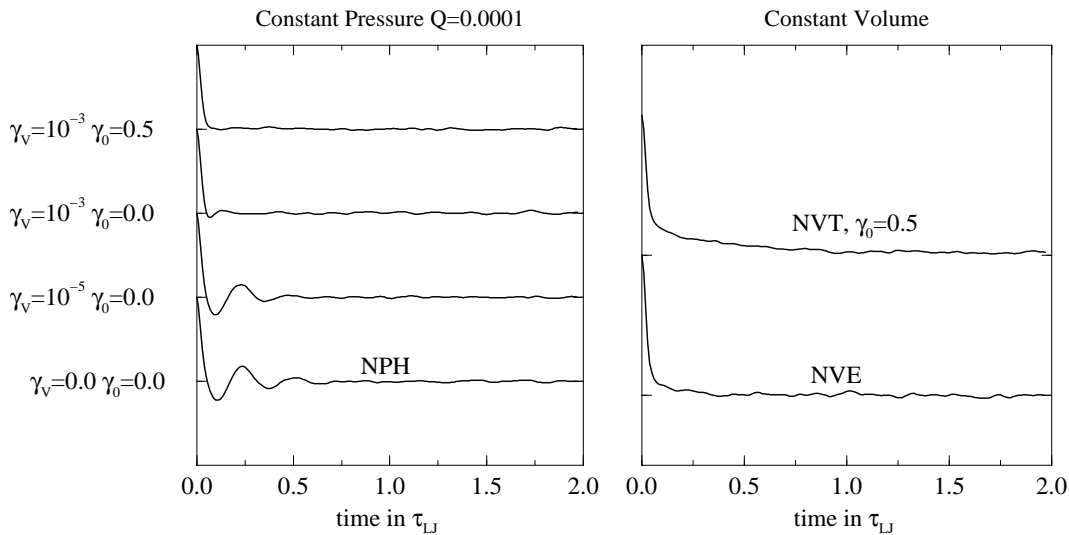,width=14cm}}
  \caption{
    Same as Fig. \protect\ref{fig:pr-autocorrelation}, but on shorter time
    scales. Even for the optimum value $Q = 0.0001$, box oscillations still
    remain, which do not occur in the pressure autocorrelation function in a
    constant volume simulation (right hand side). The use of a friction
    $\gamma_V$ damps out the oscillations, and a combination of $\gamma_V =
    0.001$ with $\gamma_0 = 0.5$ leads to the best agreement with the NVE
    result.  }
  \label{fig:pr-high-res}
\end{figure}

Figure \ref{fig:pr-autocorrelation} illustrates the problem of the ``ringing''
in this particular system, by displaying the autocorrelation function of the
pressure fluctuations for various simulation parameters. The simulated
temperature is $k_B T = 1.0$, while the pressure was fixed at the rather low
value $P = 1.0$. The mean volume for the 100 particles is $V = 262.7$,
corresponding to a density $\rho = 0.38$. The compressibility at this state
point is $\kappa_{T} = 0.3$, such that for a box mass of $Q = 0.1$ one finds
$\Omega_0 \approx 0.36$ for the ringing frequency or $2 \pi / \Omega_0
\approx 18$ for the oscillation time. The figure shows that the box indeed
oscillates, and that the frequency of the oscillations has been estimated
correctly. The left part of the figure is for pure undamped Andersen dynamics
where the dependence on $Q$ is shown. Choosing a value of $Q=0.0001$ leads to
a very fast relaxation of the pressure autocorrelation function. The
theoretical prediction for the best box mass for the molecular frequency of
$\omega_0 \approx 8.5$ in this system leads to $Q_{opt} = 1 / (V \kappa_T
\omega_0^2) \approx 0.00018$. But as seen in Fig. \ref{fig:pr-high-res},
even for this value of $Q$ the oscillations still remain on a very short time
scale. Only a value of $\gamma_{V} = 0.001$, close to $\gamma_{Vc} = 2 Q
\Omega_0 \approx 0.002$ suppresses the fluctuations efficiently and the
autocorrelation function resembles the autocorrelation function in the NVE
ensemble. Interestingly, it is also seen that for constant volume the pressure
relaxation is considerably slower if the molecular damping is turned on.
Conversely, this behavior is practically absent in the constant pressure case,
where actually the ``best'' autocorrelation function was found for $\gamma_{V}
= 0.001$ (as discussed), combined with some additional molecular friction
$\gamma_{0} = 0.5$.

The statistical accuracy of an observable is given by the ratio between
simulation time and the integrated autocorrelation time $\tau$, i.~e. the
value of the time integral over the normalized autocorrelation function
\cite{BinderHMK}. From that perspective, a slow decay with many oscillations
is actually not particularly harmful, since the integral value is rather
small, due to cancellation. However, the result of Ref. \cite{BinderHMK} holds
only in the asymptotic limit where the simulation time is substantially longer
than the decay of the correlation function. Moreover, the numerical
integration of an oscillatory function converges only slowly and is hence
rather awkward. For these reasons, a simulation algorithm which avoids
oscillations is clearly preferable. In order to illustrate this further,
Table \ref{tab:int-autocorr} lists $\tau$ for the autocorrelation functions
shown in Fig. \ref{fig:pr-high-res}. The smallest $\tau$ is actually found
for the pure Andersen NPH simulation. However, turning on the damping
increases $\tau$ only by less than a factor of two, while the decay
time decreases nearly by a factor of ten.

The case of a box whose mass has, for reasons of separation of time scales,
been chosen substantially larger than the optimum value, is illustrated in the
upper right part of Fig. \ref{fig:pr-autocorrelation} ($Q = 0.1$, i.~e. three
orders of magnitude larger than $Q_{opt}$). The autocorrelation function
decays most rapidly for $\gamma_{V} = 0.1$, as theoretically expected
($\gamma_{Vc} = 0.072$). Compared to undamped dynamics at the same mass, this
is a considerable improvement. Nevertheless, this decay is still substantially
slower than what one can obtain if also the mass is chosen optimally (Fig.
\ref{fig:pr-high-res}) --- this is simply the price which is being paid for
achieving realistic molecular motion. For practical purposes, the decay
obtained for $Q = 0.1$, $\gamma_{V} = 0.1$ is quite acceptable.

\section{Conclusion}
\label{sec:conclusion}

We have discussed the algorithm of Andersen \cite{Andersen:80.1} and the
generalization to stochastic piston motion by Feller {\em et al.}
\cite{Feller:95.1}, generalizing it even further to also
include stochastic motion of the particles. We gave a straightforward
proof that the NPT ensemble is produced. The implementation by means
of a symplectic algorithm is particularly stable and well--suited 
for MD problems. Another important point, the choice of the right
simulation parameters, was studied both theoretically and
numerically, and a guideline for their optimum values was given.
We view this algorithm as a particularly useful realization of
the constant--pressure ensemble.

\section*{Acknowledgments}

Fruitful and critical discussions with Patrick Ahlrichs, Markus
Deserno, Alex Bunker and Kristian M\"uller--Nedebock are gratefully
acknowledged.

\bibliographystyle{prsty}

\begin{table}[htbp]
  \begin{center}
    \begin{tabular}[t]{|c|c|c|c|}
      $Q$      &  $\gamma_V$  & $\gamma_0$ & $\tau$ \\
      \hline
      NVE      &  NVE         &  0.0       & 0.042  \\
      NVT      &  NVT         &  0.5       & 0.127  \\      
      0.0001   &  0.0         &  0.0       & 0.018  \\
      0.0001   &  $10^{-5}$   &  0.0       & 0.023  \\
      0.0001   &  $10^{-3}$   &  0.0       & 0.03   \\
      0.0001   &  $10^{-3}$   &  0.5       & 0.031 
    \end{tabular}
    \caption{
      Integrated autocorrelation time $\tau$ for the parameter combinations of
      Fig. \protect\ref{fig:pr-high-res}.  }
    \label{tab:int-autocorr}
  \end{center}
\end{table}

\end{document}